\documentstyle[aps,preprint]{revtex}
\begin{document}
\draft
\title{Small excitonic complexes in a disk-shaped quantum dot}
\author{Ricardo P\'erez$^{1,2}$\cite{ricardo} and 
 Augusto Gonzalez$^1$\cite{augusto}}
\address{$^1$Centro de Matem\'aticas y F\'\i sica 
 Te\'orica, Calle E No. 309, Ciudad Habana, Cuba\\
 $^2$Universidad de Antioquia,
 A.A. 1226, Medell\'\i n, Colombia}
\date{\today}
\maketitle

\begin{abstract}
Confined excitonic complexes in two dimensions, consisting of
$N_e$ electrons and $N_h$ holes, are studied by means of
Bethe-Goldstone equations. Systems with up to twelve pairs,
and asymmetric configurations with $N_e\ne N_h$ are considered.
The weak confinement regime gives indication of weak binding 
or even unbinding in the triexciton 
and the four-exciton system, and binding in the higher complexes.  
\end{abstract}


\section{Introduction}
Although the excitonic matter as a research object has already a
relatively long history, dating back to pioneer theoretical ideas in
the sixties \cite{Keldysh}, the study of small excitonic complexes
confined in regions of nanometer scale became possible only recently,
and is closely related to dramatic advances in semiconductor
technology.

Due to the smallness of the exciton lifetime in semiconductors 
(hundreds of picoseconds, sometimes even greater), signatures of
excitonic complexes are looked for mainly in optical experiments.
Exciton and biexciton peaks in photoluminescence have been 
identified in quantum dots \cite{qdots1,qdots2}. 
Recently, a confocal microscopy technique has
been applied to resolve the luminescence coming from a single
self-assembled quantum dot \cite{exp2}. The power of the excitation 
laser is such that up to six pairs are excited simultaneously in
the dot. Clearly distinct lines coming from transitions between 
multiexcitonic complexes are identified.

The calculations presented in our paper, although unrealistic, 
are inspired by the experimental work \cite{exp2}. We address a
different question, i. e. the dependence of the ground-state 
energy on the number of pairs. From this dependence, the lowest 
optical absorption line related to the creation of an 
electron-hole pair in the background of $N$ 
photo-excited pairs may be extracted.
   
On the other hand, small excitonic systems have
been widely studied theoretically in order to describe
the optical properties of bulk and low-dimensional semiconductors. 
To the best of our
knowledge, one of the most complete studies is the paper \cite{varga},
in which a variational approach is used to compute the properties
of two- and tridimensional free (not confined) excitons with up to
five particles. Our work has also a close connection with that paper.

By using the Bethe-Goldstone (BG) equation to compute the 
ground-state energies,
we are able to rise the number of particles in the complex from a
maximum of five in the variational approach \cite{varga} to 24. 
Asymmetric systems with $N_e\ne N_h$ are studied as well. 
In the weak confinement regime, we can also address the question 
about the stability of the free complexes. And in fact, our results
give indications that the free triexciton and the four-exciton system 
are unstable (or weakly bound),
whereas the higher complexes are stable. These properties have a 
distinct trace
in the optical absorption, and may be confirmed by the
experiments.   

The present paper is complementary to \cite{bcs}, in which a BCS 
variational
approach is used to study intermediate-size complexes, containing from
12 to 180 particles. For those sizes where both approaches overlap, as
a rule BCS gives lower energy values for strong confinement, whereas BG 
gives better energies in the weak confinement regime.

\section{The Bethe-Goldstone equation}

We will study a two-dimensional model of $N_e$ electrons and $N_h$ holes
confined by a parabolic potential in the plane of motion, and 
interacting 
{\it via} pure Coulomb potentials. Only one conduction and one valence 
band, both ideally parabolic, will be considered, and the masses of holes
and electrons will be supposed to be equal. 
By choosing the scales of distances and energies as 
$\sqrt{\hbar /m\omega _0}$ and $\hbar \omega _0$, we get the 
dimensionless Hamiltonian

\begin{equation}
H=\sum\limits_{i=1}^{N_e}\left( \frac{p_i^2}2+\frac{r_i^2}2\right)
+\sum\limits_{i=1}^{N_h}\left( \frac{p_i^2}2+\frac{r_i^2}2\right)
+\beta \left [\sum\limits_{i<j}^{N_e}\frac{1}{r_{ij}}
+\sum\limits_{i<j}^{N_h}\frac{1}{r_{ij}}
-\sum\limits_{i}^{N_e}\sum\limits_{j}^{N_h}\frac{1}{r_{ij}} \right ],
\label{hamilton}
\end{equation}

\noindent
where the $\vec{r_i}$ are in-plane 
coordinates for the particles, $r_{ij}=|\vec{r_i}-\vec{r_j}|$, 
$\omega _0$ is the frequency of the confining
potential, $\kappa $ is the dielectric constant of the semiconductor, 
$\beta =\sqrt{E_c/(\hbar \omega _0)}$, and $E_c=me^4/(\kappa^2 \hbar^2)$ 
is the characteristic Coulomb energy.

$\beta$ is the only parameter entering the Hamiltonian (\ref{hamilton}).
It may be thought of as the inverse of the confinement strength.
In the $\beta=0$ limit, we have a picture of non-interacting fermions 
which energy levels and one-particle quantum states are those of the 
$2D$ harmonic oscillator. The electron (hole) states are grouped 
into shells with 
``magic numbers'' $N_e,N_h=2,6,12,...$. As $\beta$ grows, correlations 
between particles become more and more important and the previous 
picture is modified. 

One way to go beyond the independent-particle picture is the 
independent-pair approximation in which an exact treatment of the 
two-body
correlations is made. Its main component is the Bethe-Goldstone (BG) 
equation, extensively 
used in nuclear matter and finite nucleus calculations 
\cite{b-g,feshbach}. 

The BG equation applies only to fermionic systems. It describes 
the motion of an independent pair of fermions in the system. The rest
of the particles excerpt an indirect influence on the pair motion through
the Pauli principle. The equation takes the form:

\begin{equation}
\left( T_1+T_2+Q_{\alpha \gamma }V\right) \psi _{\alpha \gamma }=
E_{\alpha
\gamma }\psi _{\alpha \gamma }, \label{B-G}
\end{equation}

\noindent
where $\alpha$ and $\gamma$ label the states of each fermion in the pair. 
These states are below the Fermi level. The $T_i$ are the one-particle 
terms in the Hamiltonian, $V$ is the two-body interaction potential 
(Coulomb), and $Q_{\alpha\gamma}$ is a projection operator given by

\begin{equation}
Q_{\alpha \gamma }=\left| \alpha \gamma \right ) \left ( \alpha
\gamma \right| +\sum\limits_{\mu' ,\lambda' }\left| \mu' \lambda' \right )
\left ( \mu' \lambda' \right|,
\end{equation}

\noindent
where the sum runs over states above the Fermi level. $Q$ projects a 
given function onto states over the Fermi level. 
$\left ( \vec{r_1} \vec{r_2} | \alpha \gamma \right )$ is the 
unsymmetrized 
product of two non-interacting one-particle eigenfunctions 
(we use the notation given in 
\cite{blaizot}). $E_{\alpha \gamma }$ is the pair energy, 
and $\psi _{\alpha \gamma }$ -- its wave function. 

Equation (\ref{B-G}) is formally
similar to a pair scattering equation, except for the presence of the 
projection operator (and the fact that all of the states in the 
external quadratic potential are bound states). 
The pair wave function $\psi _{\alpha \gamma }$ is  
looked for in the form

\begin{equation}
\psi _{\alpha \gamma }=\left| \alpha \gamma \right )
+\sum\limits_{\mu' ,\lambda' }C_{\mu' \lambda' }^{\alpha \gamma }\left| 
\mu' \lambda' \right ),
\end{equation}

\noindent
and the total energy  is computed from

\begin{equation}
E=\sum\limits_{\alpha}\epsilon _{\alpha}^{(0)}+
\sum\limits_{\alpha < \gamma} 
\epsilon _{\alpha \gamma}, 
\label{energy}
\end{equation}

\noindent
where $\epsilon _{\alpha \gamma }=E_{\alpha \gamma }-
\epsilon _{\alpha}^{(0)}
-\epsilon _{\gamma}^{(0)}$. The corrections coming from the BG equations
are proven to be equivalent to summing up 
all the ladder diagrams in the linked-cluster expansion for the 
energy \cite{day}.

Multiplying (\ref{B-G}) from the left by $\left (\mu \lambda \right|$ or
$\left (\alpha \gamma \right|$, we get

\begin{eqnarray}
(\epsilon_{\mu}^{(0)}+\epsilon_{\lambda}^{(0)}-E_{\alpha \gamma}) 
C_{\mu \lambda}^{\alpha \gamma} 
+\sum\limits_{\mu' ,\lambda' }
\left ( \mu \lambda \right |V \left| \mu' \lambda' \right )
C_{\mu' \lambda' }^{\alpha \gamma } 
=-\left ( \mu \lambda \right | V \left| \alpha \gamma \right ),
 \label{sys} \\
E_{\alpha \gamma}=\epsilon_{\alpha}^{(0)}+\epsilon_{\gamma}^{(0)}
+\left ( \alpha \gamma \right | V \left| \alpha \gamma \right )
+\sum\limits_{\mu' ,\lambda' }
\left ( \alpha \gamma \right |V \left| \mu' \lambda' \right )
C_{\mu' \lambda' }^{\alpha \gamma }. \label{tras}
\end{eqnarray}

Equation (\ref{sys}) may be seen as a linear system of equations for the
coefficients $C_{\mu \lambda }^{\alpha \gamma }$, from which we obtain
$C_{\mu \lambda }^{\alpha \gamma }=
C_{\mu \lambda }^{\alpha \gamma }(E_{\alpha \gamma })$. 
Then, the transcendental equation (\ref{tras}) is solved for  
$E_{\alpha \gamma }$.

One shall notice that the Coulomb 
interaction does not change neither the angular momentum of the pair nor 
the spin of the particles. The matrix elements are real 
and have the following properties:

\begin{equation}
\left ( \alpha \gamma \right |V \left| \mu \lambda \right )=
\left ( \gamma \alpha \right |V \left| \lambda \mu \right )=
\left ( \mu \lambda \right |V \left| \alpha \gamma \right ).
\end{equation}

\noindent
Due to these properties, the matrix entering the linear system is 
symmetric.

In order to solve the BG equation we have to identify, given an 
initial pair, all the possible final states above the Fermi level 
preserving
angular momentum and spin. Then we solve the linear system 
and the transcendental equation to get the pair energy. One shall 
take into account that there are three kinds of pairs in the system, 
namely e-e, h-h and e-h.
In the second term of (\ref{energy}), $\alpha<\gamma$ means that
for identical particles (e-e and h-h) a state $| \alpha \gamma )$
should be counted only once.
For e-h pairs, however, we should take into account the two possibilities, 
i. e. $| \alpha_e \gamma_h )$ and $| \alpha_h \gamma_e )$.

In our calculations, up to 272 harmonic-oscillator one-particle states
(16 shells) are included. Whenever possible, the e-h, spatial inversion 
and
time-reversal symmetries are used to reduce the actual number of 
equations to 
solve. For example, in the $N_e=N_h=6$ problem, 15 systems of linear 
equations with roughly 700 unknowns each, and 15 nonlinear equations for
the pair energies are solved.

\section{Symmetric ($N_e=N_h$) systems}

In this section we present the results obtained for symmetric systems, 
where 
$N_e=N_h=N$. The parameter $\beta$ is varied in the interval (0,~2.5).

The first step in our Bethe-Goldstone calculation is to define filled and
empty levels, i. e. the Fermi surface. To this end, a
Hartree-Fock calculation was implemented. The harmonic-oscillator states 
were filled in accordance with the Hartree-Fock results. We show them in
spectroscopic (nuclear) notation in Table \ref{tab1}. $2p_-$, for example,
means the second level with $l_z=-1$. $S_e$, and $S_h$ refer to the total 
electron and hole spins respectively, and $L$ is the total angular
momentum (along the $z$-axis).

\subsection{The biexciton}

For the biexciton, our starting configuration is one in which the 
first harmonic oscillator shell is filled for both electron and holes
(see Table \ref{tab1}). This state
has zero total angular momentum and total spin. 
In order to write down the Bethe-Goldstone equations we
shall identify all possible pairs below the Fermi level. In the e-e and
h-h sectors there is only one pair, but we have 4 e-h pairs. For a given
initial pair, the number of final pair states above the Fermi level 
depends on the number of harmonic oscillator shells, $N_{shell}$, 
included in the calculations. Figure 1 
shows the results for the energy as a function of $\beta$ and $N_{shell}$.
The solid line is a two-point Pad\'e approximant, which construction is 
described below. 

A few remarks shall be given at this point. First, notice that the 
convergence is slow. Second, the BG results do not exactly reproduce
the perturbative
energies at small values of $\beta$. It means simply that the 
characteristic
distances are much smaller than the characteristic pair dimensions, and 
thus the independent pair approximation breaks down. Fortunately, in this
strong-confinement regime the energy is dominated by the one-particle
energies, which are properly accounted for by the BG approach. Finally,
the BG energies are seem to overestimate the actual binding energies
at large $\beta$.

The two-point Pad\'e approximant for the ground-state energy is 
constructed 
in the following way\cite{MacD,A98}. The asymptotic expansions

\begin{equation}
\left. E\right|_{\beta\to 0}=b_0+b_1 \beta+b_2 \beta^2+{\cal O}(\beta^3)
\label{serienul}
\end{equation}

\begin{equation}
\left.E\right|_{\beta\to\infty}= a_0 \beta^2 + a_2 +
 {\cal O}(1/\beta^2),
\label{eq13}
\end{equation}

\noindent
are used, where $b_0=4$, $b_1=-2.50662$, $b_2=-2.92$ \cite{bcs}, and
$a_0=-2.1928$ \cite{varga}, $a_2=1$. $a_0$ is the 
ground-state energy of the free biexciton in two dimensions, and
$a_2$ comes from the center of mass oscillation,
contributing with a one to the ground-state energy. 

The two-point approximant is a rational function interpolating
between (\ref{serienul}) and (\ref{eq13}):

\begin{equation}
P(\beta)=b_0+b_1\beta+\frac{b_2\beta^2+p_3\beta^3+
 p_4\beta^4}{1+q_1\beta+q_2\beta^2}.
\label{eq19}
\end{equation}

\noindent
The values of the parameters are the following,
$p_3=0.525372$, $p_4=-2.06239$, $q_1=0.83554$ and $q_2=0.940527$.

\subsection{N=3, 4, 5, 6 and 12}

Next, we present results for $N=3,4,5,6$ and 12, computed with 
$N_{shell}=16$. The occupied harmonic-oscillator states are indicated
in Table \ref{tab1}. Note that, in some cases, the Hartree-Fock 
approximation 
does not suggest a unique state, thus we performed BG calculations
with different occupations, and select the one corresponding to the
minimal energy. For example, in the $N=3$ system we performed 
calculations 
also for a configuration very similar to the one given in Table 
\ref{tab1}, but with total momentum L=2.

The results are drawn in Fig. \ref{fig2} in a ``scaled'' form, i. e. 
$E_{gs}/N^{3/2}$ vs. $\beta/N^{1/4}$. This scaling comes from 
the dependence $b_0\approx 4/3~ N^{3/2}$, $b_1\approx -0.96~ N^{5/4}$,
$b_2\approx -1.65~ N$ and $a_0\approx -N$ for large $N$\cite{bcs}, but
it is nicely satisfied even for the smallest systems. In this scaled 
drawing,
the $N=4$ cluster clearly distinguishes as the less bound one.

The results at the largest values of $\beta$, i. e. in the weak
confinement regime, can be taken as indications of stability or 
instability
of the free clusters. The triexciton seems to be unbound (like in three
dimensions\cite{varga2}), and the four-exciton system -- evidently unbound.
However, the larger clusters are likely to be stable. The situation 
may be 
analogous to nuclei, where there is a small instability island around
atomic number five.

The differences $E_{N+1}-E_{N}$ are shown in Fig. \ref{fig3}. We notice
that these differences enter the expression

\begin{equation}
h\nu=E_{gap}+E_z^e+E_z^h+E_{N+1}-E_{N},
\label{frecuencia}
\end{equation}

\noindent
for the frequency of light creating a new electron-hole pair in the
background of $N$ pairs. $E_{gap}$ is the semiconductor gap, and 
$E_z^{e,h}$ are the electron (hole) confinement energies in the 
$z$-direction.

As following from Fig. \ref{fig3}, the absorption peak corresponding
to (\ref{frecuencia}) will exhibit an interesting behaviour as 
a function of the
number of photoexcited pairs, $N$. High values of $N$ can be 
reached by rising the laser excitation power. When $N$ is around four, 
we shall
observe a highly blueshifted line, followed by a redshifted one as the 
power is further increased. The conclusions coming from our 
oversimplified model have only a qualitative predictive power. The
effect is so pronounced, however, that we expect it may be 
observed in experiments.    

\section{Asymmetric cases}

We also studied non-symmetric or non-neutral systems, in which 
$N_e\ne N_h$. In particular, the cases
$N_e=4$, $N_h=2$ and $N_e=6$, $N_h=2$. The results for
their ground state energies are shown in Fig. \ref{fig4}. In a 
non-neutral system, the energy is an increasing function of 
$\beta$ for low $\beta$ values, as for electrons \cite{GPP97},
but for weaker confinement the -$\beta^2$ biexcitonic contribution
dominates over the $\beta^{2/3}$ repulsion. A biexciton plus
remaining electrons shall be seen at very large $\beta$. The 
$N_e=6$, $N_h=2$ system clearly reveals the change in slope.
 
\section{Conclusions}

In the present paper, we have shown that the Bethe-Goldstone 
equations may be used as a powerful method to study small confined
excitonic clusters. Pure electronic quantum dots may be studied
as well, and external electric and magnetic fields can be 
easily included in the calculations. Larger unbalanced systems
with only one exciton may be studied to obtain the optical 
absorption and photoluminescence of small electronic quantum dots.

Our numerical results suggest a small instability island for the 
free clusters around $N_e=N_h=4$. We showed that this instability
causes the appearance of distinct absorption peaks 
as the laser excitation power is raised. 

More realistic calculations, closely related to the experimental
results \cite{exp2}, are in progress.

\acknowledgements
The authors acknowledge support from the Universidad de Antioquia,
Medellin. Part of this work was done during a visit to the Abdus
Salam ICTP under the Associateship Scheme and the Visiting Young
Student Programme. Useful discussions with E. Lipparini are also 
acknowledged.

\begin{figure}
\caption{Convergence of the biexciton energy as a function
 of the number of harmonic-oscillator shells included in the
 calculations.}
\label{fig1}
\end{figure}

\begin{figure}
\caption{Energies of the N-exciton systems as a function of 
 $\beta$.}
\label{fig2}
\end{figure}

\begin{figure}
\caption{Energy differences $E_{N+1}-E_{N}$ for the smallest
 clusters studied.}
\label{fig3}
\end{figure}

\begin{figure}
\caption{Energies of the $N_e=4$, $N_h=2$, and $N_e=6$,
 $N_h=2$ systems.}
\label{fig4}
\end{figure}

\begin{table}
\caption{Occupied electron and hole states for the BG calculations.}
\label{table1}
\begin{tabular}{|c|c|c|c|c|c|c|}
$N_e$ & $N_h$ & electrons  & holes & $S_e$ & $S_h$ & L \\
\hline\hline
2 & 2  & $1s^2$ & $1s^2$ & 0 & 0 & 0 \\
3 & 3  & $1s^2 1p_+^1$ & $1s^2 1p_-^1$ & 1/2& 1/2 & 0 \\
4 & 4  & $1s^2 1p_+^1 1p_-^1$ & $1s^2 1p_+^1 1p_-^1$ & 1 & 1 & 0 \\
5 & 5  & $1s^2 1p_+^2 1p_-^1$ & $1s^2 1p_+^1 1p_-^2$ & 1/2 & 1/2 & 0 \\
6 & 6  & $1s^2 1p_+^2 1p_-^2$ & $1s^2 1p_+^2 1p_-^2$ & 0 & 0 & 0 \\
12&12  & $1s^2 1p_+^2 1p_-^2 2s^2 1d_+^2 1d_-^2$ & $1s^2 1p_+^2 
1p_-^2 2s^2 1d_+^2 1d_-^2$  & 0 & 0 & 0 \\
4 & 2 &  $1s^2 1p_+^1 1p_-^1$ & $1s^2$ & 1 & 0 & 0 \\
6 & 2 &  $1s^2 1p_+^2 1p_-^2$ & $1s^2$ & 0 & 0 & 0 \\
\end{tabular}
\label{tab1}
\end{table}

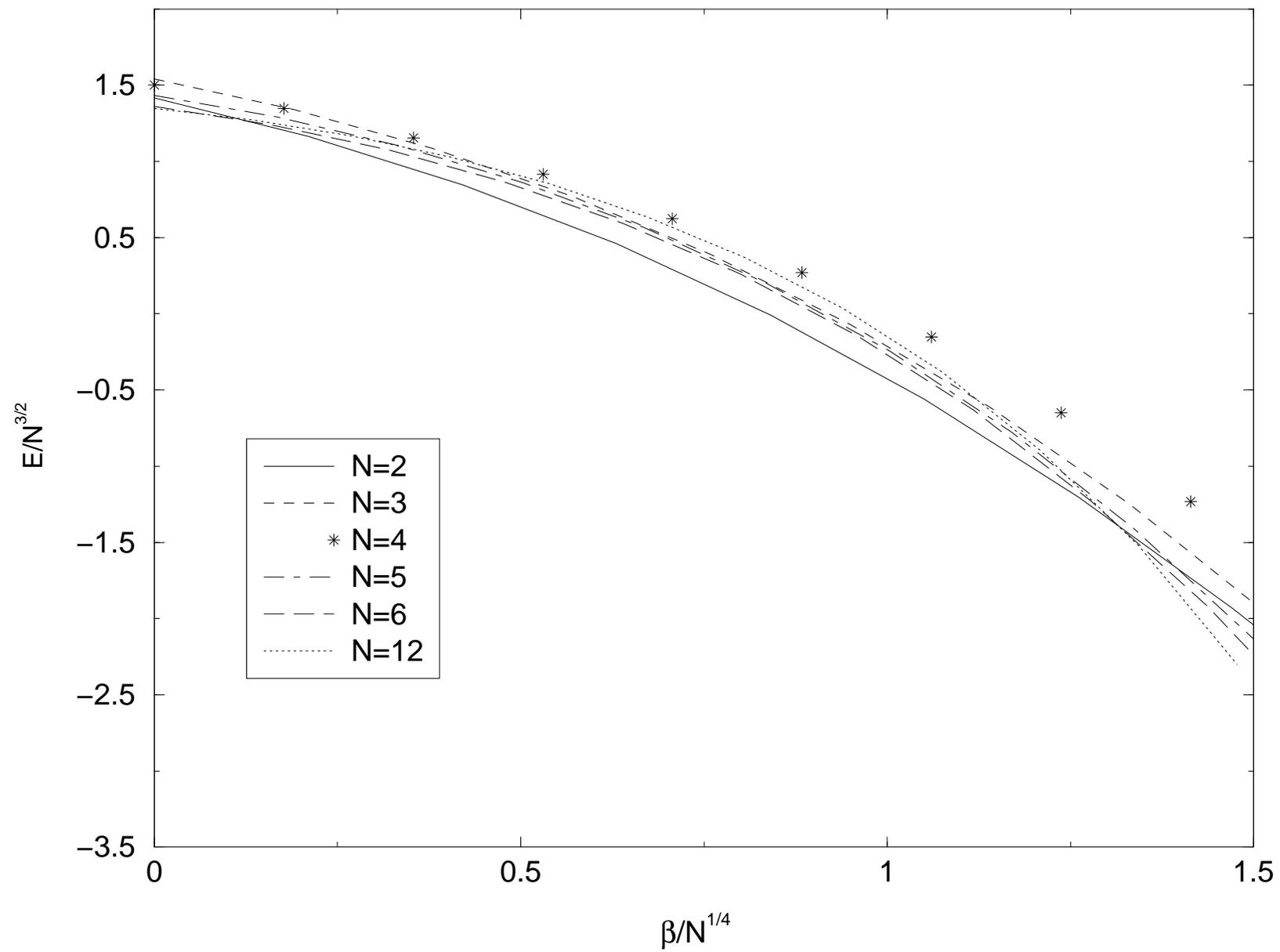
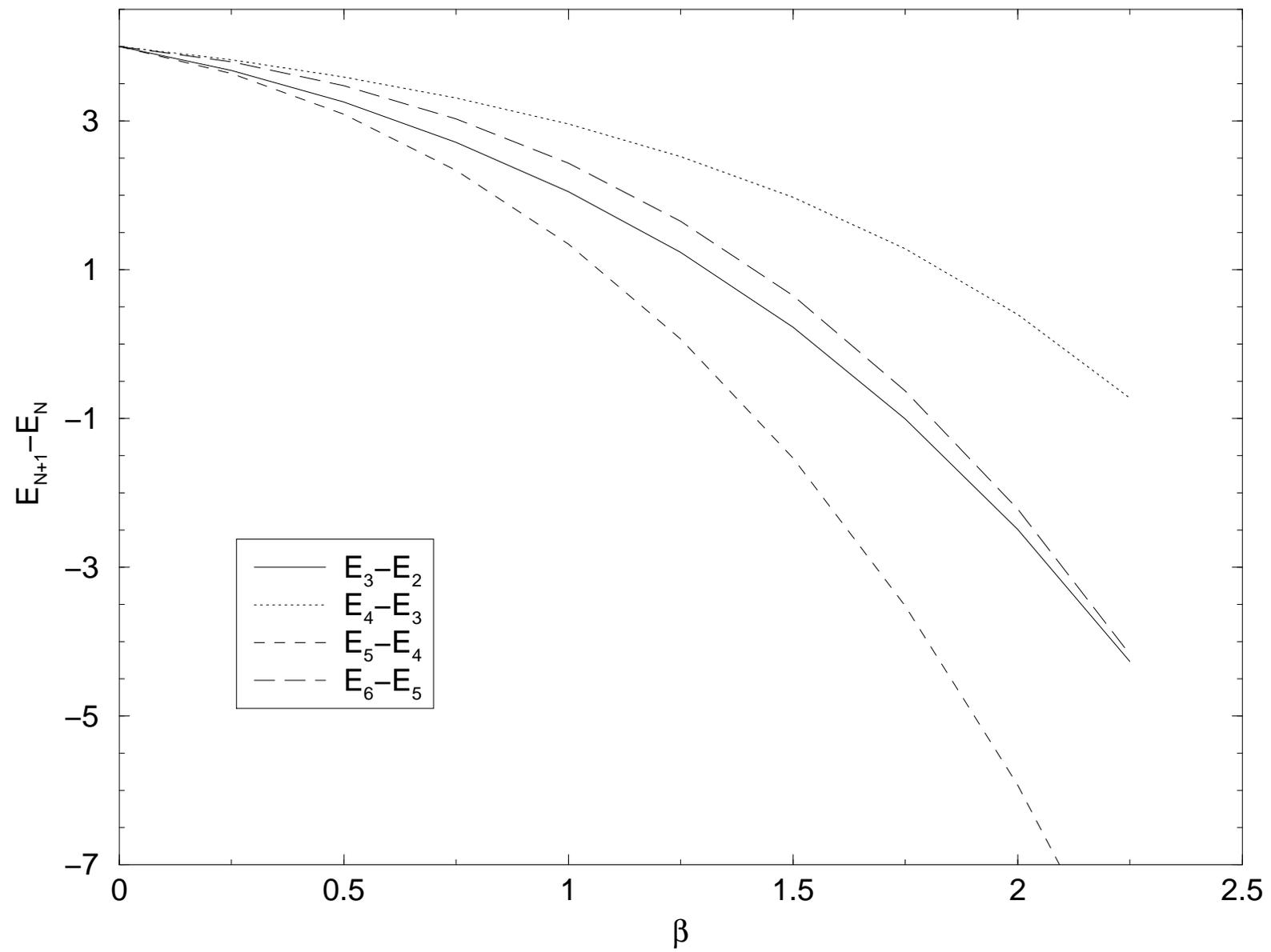
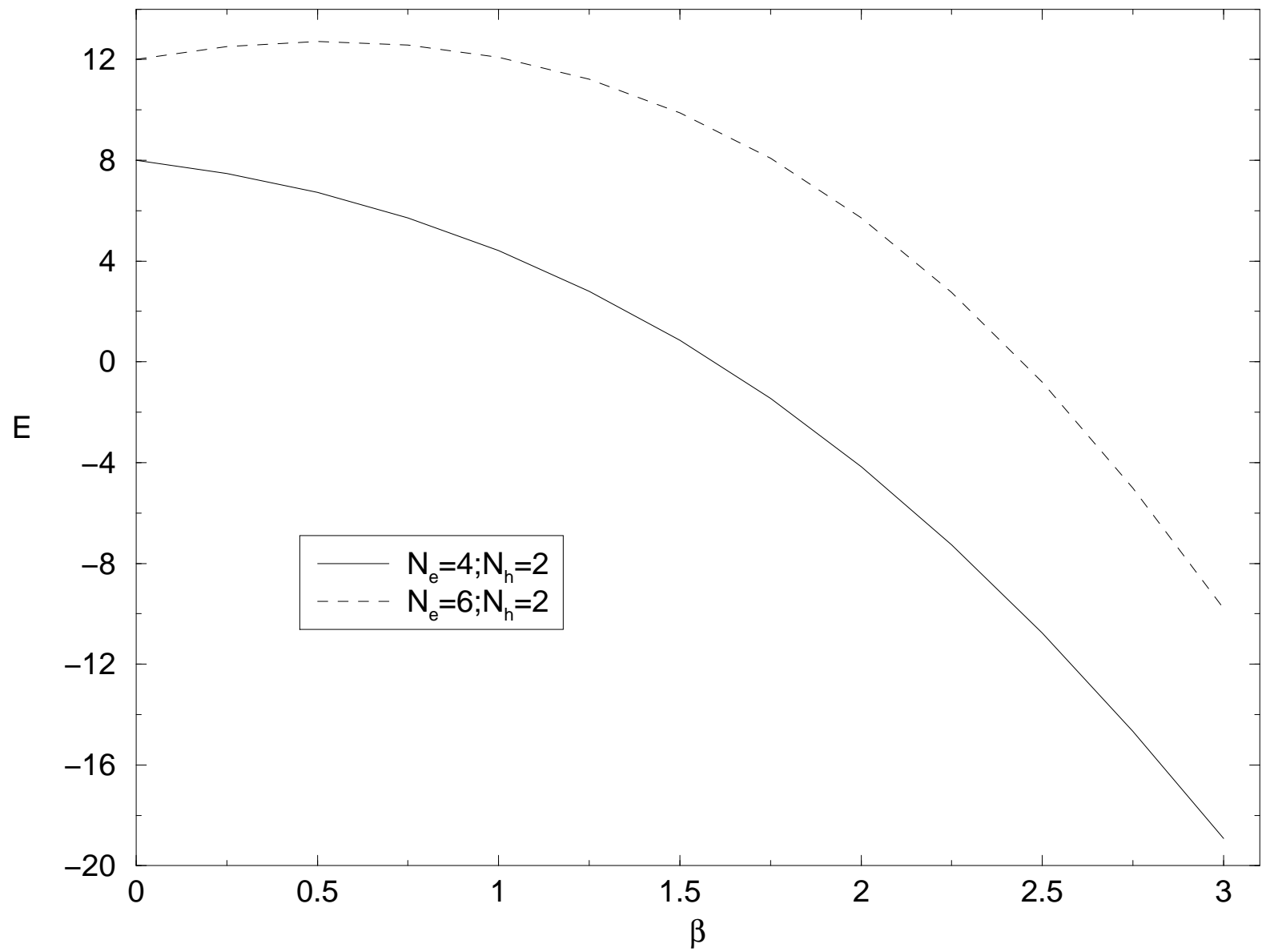
\end{document}